# Smart Metadata in Action: The Social Impact Data Commons

Joanna Schroeder[1,2], Alan Wang[2], Kathryn Linehan[2], Joel Thurston[2], and Aaron Schroeder[2]

[1] University of Virginia, Charlottesville, Virginia, United States
[2] Drexel University, Philadelphia, Pennsylvania, United States

**Abstract**
This article describes the use of metadata and standards in the Social Impact Data Commons to expose official statisticians to an innovative project built on actionable and evaluable metadata, which produces a FAIR data system. We begin by introducing the concept of the Data Commons, focusing on its features, and presenting an overview of current implementations of the Data Commons. We then present the core metadata case study, demonstrating how smart metadata support the Data Commons. We also present evaluations of our core metadata, including its adherence to the FAIR guidelines. We conclude with a discussion on our future metadata and standards-related projects to support the Social Impact Data Commons.

## 1. Introduction

### 1.1. What is a Data Commons?

Local stakeholders often lack the in-house expertise to meet their specific data needs. To address local data needs, we propose development of the Data Commons. The Data Commons as a collaborative and open knowledge repository. To build the Data Commons, we implement the Community Learning through Data Driven Discovery (CLD3) framework [1].
Figure **1** supplies an overview of CLD3, which emphasizes stakeholder collaboration and centers stakeholder data needs and user preferences at the core of every project. CLD3 lays the groundwork for arriving at specific and meaningful local insights characteristic of the Data Commons.

**Figure 1:** The Community Learning through Data Driven Discovery Framework

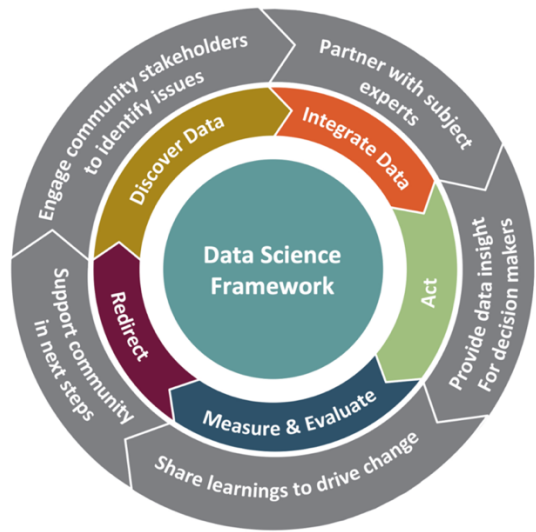

The Data Commons, then, necessarily co-locates data to address a variety of stakeholders and user profiles. The Data Commons includes not only a range of datasets, but also data products (dashboards and APIs), data tools (statistical packages), and data stories (narratives that bring data to life). Importantly, the Data Commons is created using lightweight infrastructure, taking advantage of the features of open-source tools like serverless hosting on GitHub. A transdisciplinary research team at the Social and Decision Analytics Division of the University of Virginia's Biocomplexity Institute created, updates, and maintains the Data Commons.

The Data Commons infrastructure has supported three implementations thus far.
- We produced the Social Impact Data Commons for county officials in the National Capital Region of the United States. The National Capital Region includes Washington, D.C. and 24 counties and independent cities surrounding it in Virginia and Maryland. The Social Impact Data Commons features data on topics relevant to local decision-makers, such as broadband access and food insecurity.
- We produced the Virginia Department of Health Rural Health Site for the Office of Rural Health within the Virginia Department of Health (VDH). The Rural Health Site contains metrics of health service accessibility to highlight rural health disparities. This implementation is now being used by the Office of Health Equity at VDH to support their Health Opportunity Index. Virginia Cooperative Extension is also a stakeholder and user for this implementation.
- We are developing **the Fairfax Women and Girls Study Data Commons** for county officials in Fairfax County, Virginia. This implementation will focus on issues of local relevance to women and girls, highlighting equity and intersectional identities.

Because they are built on a shared infrastructure, implementations of the Data Commons share common elements, such as base geographic layers and user interface components. Implementations of the Data Commons are tailored to reflect the specific data needs of stakeholder and user groups. In this paper, we will discuss metadata and standards as they relate to the Data Commons infrastructure overall and share specific examples from the Social Impact Data Commons where relevant.

## 1.2. Evaluating the Data Commons Metadata Standards

We leverage standards and metadata to support the creation of the Data Commons. Throughout our process of creating standards and metadata, we prioritize actionable metadata and evaluable metadata. In this context, we define actionable metadata as metadata whose machine-actionable design allows it to extend beyond description of data. We define evaluable metadata as metadata that can measured and benchmarked. By favoring actionable and evaluable metadata, our research team can rely more heavily on machine-automated processes to monitor the successes and failures of the current system. Thus far, we have evaluated metadata based on applicable domain standards as well as an internal global testing process.

Our process of developing and evaluating the Data Commons core metadata and standards, as well as the project overall, is ongoing. Because the Data Commons necessarily involves the confluence of a wide variety of data, as well as a variety of stakeholders and researchers, we have found that the importance of codifying and disseminating standards is highly important. Moreover, automating the system also relies on adhering to robust standards. We began the Data Commons project with no in-house metadata experts on the team, and we quickly recognized the need to develop this expertise. Since then, we have instituted a metadata team for the project. The team meets weekly to discuss rotating topics. Occasionally the team will bring in a data producer (i.e. a project member that creates or imports a particular dataset) to collaborate with them to create better metadata or test out new techniques. The metadata team is the ultimate authority on metadata and standards for the project. For example, the team decides when to include new core metadata elements and retire old ones. As a team, we strive to continuously critique our system and improve our solutions.

We evaluate our evolving metadata and standards against applicable domain standards, specifically, the FAIR guidelines. Wilkinson et al. (2016) formally introduced the FAIR Data Principles of findability, accessibility, interoperability, and reusability [2]. They argue that building up FAIR data infrastructure is necessary for data reusability. In addition to the foundational principles, Wilkinson et al. introduce a set of guiding principles to evaluate progress towards FAIR in greater detail [2]. We acknowledge that the FAIR guidelines play a role within larger systematic changes necessary to meet ideals of open, reproducible science [3]. In Figure 2 we reproduce the FAIR Guiding Principles to reference throughout this evaluation as a benchmark to measure FAIR compliance.

**Figure 2:** The FAIR Guiding Principles (Reproduced from Wilkinson et al. (2016, p. 4)) [2]

> To be Findable:
> F1. (meta)data are assigned a globally unique and persistent identifier
> F2. data are described with rich metadata (defined by R1 below)
> F3. metadata clearly and explicitly include the identifier of the data it describes
> F4. (meta)data are registered or indexed in a searchable resource
>
> To be Accessible:
> A1. (meta)data are retrievable by their identifier using a standardized communications protocol
> A1.1 the protocol is open, free, and universally implementable
> A1.2 the protocol allows for an authentication and authorization procedure, where necessary
> A2. metadata are accessible, even when the data are no longer available
>
> To be Interoperable:
> I1. (meta)data use a formal, accessible, shared, and broadly applicable language for knowledge representation.
> I2. (meta)data use vocabularies that follow FAIR principles
> I3. (meta)data include qualified references to other (meta)data
>
> To be Reusable:
> R1. meta(data) are richly described with a plurality of accurate and relevant attributes
> R1.1. (meta)data are released with a clear and accessible data usage license
> R1.2. (meta)data are associated with detailed provenance
> R1.3. (meta)data meet domain-relevant community standards

In addition to evaluating using the FAIR principles, we have developed a testing infrastructure for evaluating the entire Data Commons system including metadata and standards. Metadata is leveraged by most tests. The testing infrastructure uses a GitHub runner. Each day the GitHub runner clones (refreshes) all linked data repositories. Next the runner executes specified tests, which are written as python scripts, on the data repositories. The results of the test are stored as HTML files and deployed as a website. Depending on the goal of the test, a test may iterate over different objects in the data repository. This includes folders or files of a certain type (e.g., JSON,CSV). Tests are developed and maintained by the Data Commons metadata team. The metadata team develops tests when a need to evaluate adherence to standards or provide feedback to data producers is identified. We use tests to measure adherence to standards and manage enforcement of metadata standards. Enforcement is balanced to align with the need to provide high-quality data products to our stakeholders. Data producers retain some latitude to maintain their own autonomy in the data production process, mostly within development data repositories. Table 1 describes the current implemented tests.

**Table 1**
**Overview of Implemented Tests**

| Test ID | Test Name | Test Description |
|---|---|---|
| T2 | /test_percent_data.html | Checks if percent data is in the range 0-100 (not 0-1) |
| T3 | /test_measure_info_structure.html | Checks whether measure info files have and only have a prescribed list of allowable keys |
| T4 | /test_measure_type.html | Checks whether measure types in data files are valid |
| T5 | /test_measure_info_missing_measures.html | Checks whether measure info files are missing any measures contained in corresponding data files |

| Test ID | Test Name | Test Description |
| --- | --- | --- |
| T6 | /test_columns.html | Checks whether or not csvs have the predetermined column names for each csv |
| T7 | /test_measure_info_keys.html | Checks whether measure info files have valid keys for each variable |
| T8 | /test_jsons.html | Checks whether encountered jsons are valid jsons that can be read |
| T9 | /test_region_type.html | Checks whether region types in data files are valid |
| T10 | /test_known_measures.html | Checks whether measures in files matches a set of known measures |
| T11 | /test_file_name.html | Checks file name conventions |
| T12 | /test_code_exists.html | Checks if distribution code exists when distribution data exists |
| T13 | /test_file_name_len.html | Checks whether or not file names are less than a prescribed 100 character limit for accessibility on windows platforms |
| T14 | /test_measure_info_extra_measures.html | Checks whether measure info files have any extra measures not contained in any corresponding data files |

## 2. Case Study: Core Metadata

### 2.1. Defining Core Metadata and Describing its Use in the Data Commons

Core metadata is an integral piece of FAIR datasets, particularly for endeavors that strive to co-locate massive amounts of data. Metadata is necessary to support adherence to each of the FAIR principles [2]. Poole (2015) argues that, especially as the landscape of science data has evolved, proper data curation is paramount [4]. Metadata is an integral piece of the data curation process [4]. Core metadata is usually defined as a minimum number of elements to maintain linkages between objects and preserve objects over time [5]. We can think of core metadata as a minimum set of elements adopted by an institution in the pursuit of FAIR data curation.

**Core metadata for the Data Commons includes the minimum set of elements to describe datasets accurately and richly, as well as support the creation of the dashboard. Our core metadata includes 17 elements, none of which we explicitly require.**
Table 2 describes the core metadata elements for the Data Commons. The main functions of our core metadata are to describe and display data. In our selection and development of core elements, we chose elements to prioritize complete and detailed descriptions of data. For example, the aggregation_method, unit, and data_type fields were recently conceptually disaggregated from a single type field to more accurately and richly describe the data. We also aim to use metadata to inform the reusability of datasets. The long_description field is critical for data producers to give a thorough account of their process of creating a dataset, thus facilitating future reuse of the data.

**Table 2**
**Core Metadata Elements for the Data Commons**

| Element | Description | Derived From | Example |
|---|---|---|---|
| aggregation_method | Identifies the type of aggregation used to combine related categories, usually within a common branch of a hierarchy, to provide information at a broader level than the level at which detailed observations are taken. (From: The OECD Glossary of Statistical Terms) | AggregationMethod (DDI/OECD) | "percent" |
| category | Thematic category or categories for the measure | Internal controlled vocabulary | "Broadband" |
| citations | Citations for methods used to produce the measure | BibTeX (LaTeX) | "lou04" |
| data_type | Type of the data, automatically inferred | DataType (DDI/W3C) | "decimal" |
| equity_category | Describes how the measure relates to equity | Internal controlled vocabulary | "Accessibility" |
| layer | Related shapefile of points to be displayed with the measure | Internal controlled syntax | "source": "https://raw.githubusercontent.com/uva-bi-sdad/sdc.health/main/Health%20Care%20Services/Urgent%20Care%20Centers/Service%20Access%20Scores/data/distribution/urgent_care_points_2022.geojson" |
| long_description | Long account of the production of the measure (no character limit). Should include detail on methods, multiple data sources, the way things were combined, decisions made to create the measure. | NA (Free text) | "Percentage of the households self-reported to not have a computer or device at home. Based on American Community Survey Table B28001 in ACS 2015/2019 5-year estimates." |
| long_name | Human readable long name for the measure (55 char limit) | NA (Free text) | "Percentage of households without a computer" |
| measure_type | Type of the measure | Internal controlled vocabulary | "percent" |
| short_description | Short account of the production of the measure (100 char limit) | NA (Free text) | "Percentage of households self-reported to not have a computer or device at home" |
| short_name | Human readable short name for the measure (40 char limit) | NA (Free text) | "Households without a computer" |
| sources | Data source from which the measure is derived | Internal controlled syntax | "date_accessed": "2022", "location": "Table B28001 ", "name": "American Community Survey", "url": "https://www.census.gov/programs-surveys/acs.html" |

| Element | Description | Derived From | Example |
|---|---|---|---|
| statement | Dynamic statement to describe the value of the data for a given geoid and year combination | Internal controlled syntax | "{value} of households in {region.name} do not have a computer or device." |
| unit | Describes the entity being analyzed in the study or in the variable, i.e. Units associated with the measure | AnalysisUnit (DDI) | "household" |
| categories | Used to populate dynamic metadata files | NA (Free text) | "NAICS72" |
| variants | Used to populate dynamic metadata files | NA (Free text) | "entry_rate" |

Dataset producers and the metadata team collect or create core metadata for every dataset in the Data Commons. The Data Commons contains a mixture of datasets that have undergone various levels of transformation by researchers. Our process of collecting or creating core metadata varies depending on the source and relevant statistical transformations the data has undergone. Many datasets that are already described with rich metadata can be manually crosswalked into our metadata standard. We try not to alter the metadata for external datasets as much as possible to preserve their integrity. For more complex datasets, the data producer collaborates with the metadata team to create accurate and rich metadata.

Our core metadata is disseminated in a JSON file (i.e. measure_info.json or measure_info) on GitHub to facilitate collaboration and versioning. GitHub leverages Git version control software. One of the advantages of version control is that our researchers and metadata team can easily collaborate on metadata. GitHub also versions repositories, keeping a historic record of files and datasets. We use the JSON file format to store our metadata. Broadly, the advantages of JSON over XML include performance, simplicity, and flexibility [6, 7]. The choice between JSON and XML is project dependent, though. In our case, we chose JSON to align with our priorities to ensure human-readability and compatibility with web interfaces.

The metadata team develops and maintains a website to efficiently disseminate metadata information to dataset producers. The decisions made by the metadata team live in the metadata GitHub repository. This repository includes codified standards, such as the set of core elements, and vocabularies. We use the information from the repository to compile the metadata website. This website is designed to be referenced by data producers and project stakeholders. Dataset producers use the website as a guide for adhering to metadata standards. Stakeholders and users use the website by searching the data dictionary to understand the data beyond what is available on the dashboard.

### 2.2. The Active Role of Core Metadata in the Data Commons

Core metadata is a key driver of the process to create the Data Commons dashboard. Our core metadata includes several elements that are used as direct inputs into our Data Commons dashboard. These elements include:
- category
- citations

- layer
- long_description
- long_name
- measure_type
- short_description
- short_name
- sources
- statement

Figure 3 annotates a screenshot of the Social Impact Data Commons with core metadata inputs.

**Figure 3**: A screenshot of the Social Impact Data Commons with core metadata features annotated

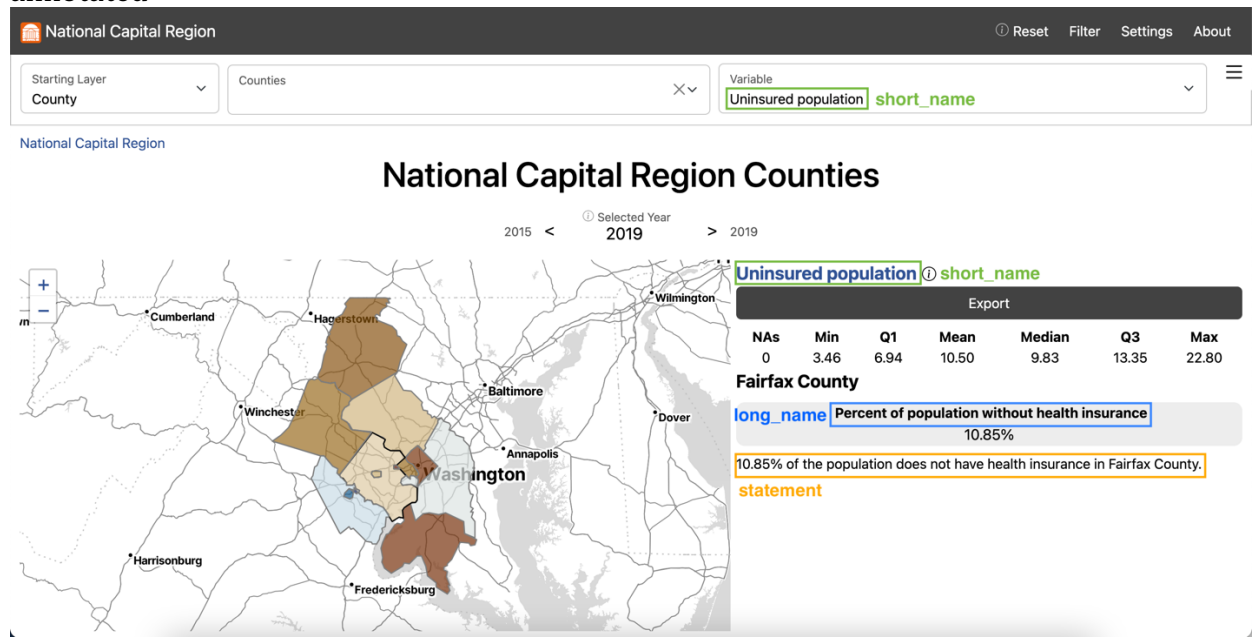

Core metadata supports the creation and dissemination of statistical products. As previously mentioned, the Data Commons is designed to co-locate datasets for stakeholders. This includes pairing of off-the-shelf statistical products with internally produced measures of local interest. Metadata directly supports how these measures are harmonized on the Data Commons dashboard. For example, stakeholders may be interested in issues of equity, such as broadband access within a county. Using the Data Commons infrastructure, we pair measures from the American Community Survey, like percentage of households reporting broadband access and neighborhood demographics, with opportunity data like internet speed and cost of broadband internet as a percentage of income retrieved from other sources. In this way, we leverage statistical products to tell meaningful data stories and target local interests.

Core metadata and metadata tools relieve burden on data producers. We do not have metadata experts contributing to the Data Commons and many researchers have no formal metadata training outside of training received on the project. As such, it is critical for us to provide our researchers with adequate training and standards to succeed in producing accurate and rich metadata. Dynamic metadata is an example of a tool designed to relieve researcher documentation burden. Instead of requiring researchers to create metadata for multiple combinations (e.g. 5 measures each for 19 industries), we train them to create a dynamic metadata file and functionally produce the final metadata. Providing researchers with the time and resources to own the production of their own metadata is an ongoing process within the Data Commons project.

### 2.3. Evaluating Core Metadata

#### 2.3.1. FAIR Evaluation

Core metadata supports the usability of the Data Commons datasets to achieve a FAIR system. Table 3 shows our self-evaluation of our progress towards adhering to the FAIR principles. We consulted Wilkinson et al. (2016) as well as interpretations and implementation considerations of FAIR principles from Jacobsen et al. (2020) to evaluate our progress towards each FAIR guiding principle [8]. We defined three categories for progress:
- Achieving: Our system meets all or nearly all of the guidance
- Working Towards: Our system implements some of the guidance
- Not Addressing: Our system meets none or very little of the guidance

We find that we begun to address most of the principles with the implementation of our metadata systems. Our system is strongest in the principles of findability and accessibility. Because our infrastructure relies so heavily on GitHub, it is in many ways necessarily findable and accessible. Our system is weakest in the interoperability principle. As shown in Table 3, only 4/17 core metadata elements are derived from a widely-adopted metadata schema. Because of this, our metadata is not easily interoperable with other schemas.

**Table 3**
**Self-Evaluation of Progress Towards FAIR Using Jacobsen et al. (2020) Guidance**

| FAIR Guiding Principle | Achieving | Working Towards | Not Addressing |
|---|---|---|---|
| F1. (meta)data are assigned a globally unique and persistent identifier | | X | |
| F2. data are described with rich metadata (defined by R1 below) | | X | |
| F3. metadata clearly and explicitly include the identifier of the data it describes | X | | |
| F4. (meta)data are registered or indexed in a searchable resource | | X | |
| A1. (meta)data are retrievable by their identifier using a standardized communications protocol | X | | |
| A1.1 the protocol is open, free, and universally implementable | X | | |
| A1.2 the protocol allows for an authentication and authorization procedure, where necessary | X | | |

| FAIR Guiding Principle | Achieving | Working Towards | Not Addressing |
|---|---|---|---|
| A2. metadata are accessible, even when the data are no longer available | | | X |
| I1. (meta)data use a formal, accessible, shared, and broadly applicable language for knowledge representation. | | X | |
| I2. (meta)data use vocabularies that follow FAIR principles | | X | |
| I3. (meta)data include qualified references to other (meta)data | | | X |
| R1. meta(data) are richly described with a plurality of accurate and relevant attributes | | X | |
| R1.1. (meta)data are released with a clear and accessible data usage license | | X | |
| R1.2. (meta)data are associated with detailed provenance | X | | |
| R1.3. (meta)data meet domain-relevant community standards | | X | |

We also self-evaluate our adherence to FAIR using the RDA FAIR Data Maturity Model [9]. Figure 4 shows the results of our maturity self-evaluation. Again, we find that our system is weakest in adherence towards interoperability, with several indicators not being considered. Our adherence towards reusability is also relatively weak. The FAIR Data Maturity Model distinguishes also the importance of indicators and we find that there is one Essential indicator we have not considered: Metadata is guaranteed to remain available after data is no longer available [9]. Addressing this shortfall should be a high-priority in our metadata strategy moving forward.

**Figure 4:** Self-Evaluation of Progress Towards FAIR using RDA FAIR Data Maturity Model Working Group (2020) Guidance
Note: Full reproduction of the FAIR Data Maturity Model Indicators available in Table A1

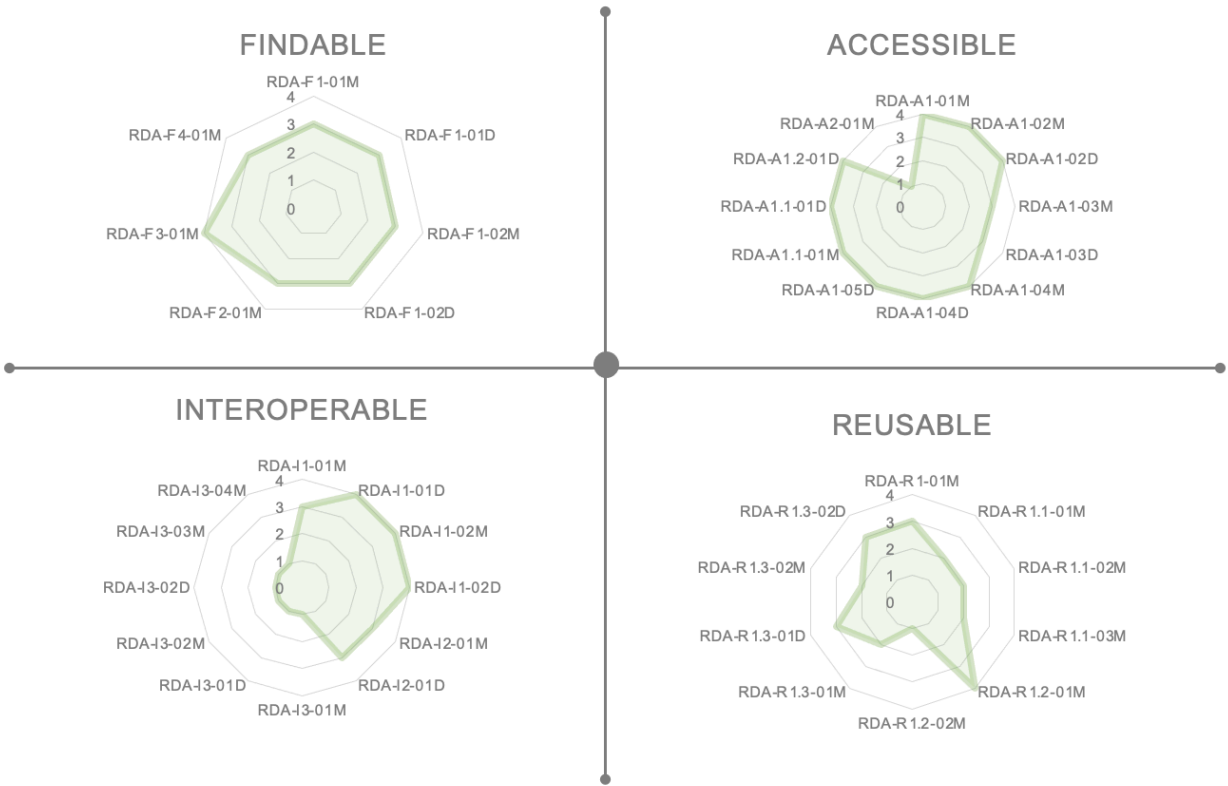

### 2.3.2. Testing Scheme Evaluation

We also evaluate core metadata through our testing scheme (see Table 1). Currently we implement five tests that address the evaluation of core metadata.

- T3: test_measure_info_structure
    - Checks whether measure_info files have and only have a prescribed list of allowable keys
- T5: test_measure_info_missing_measures
    - Checks whether measure_info files are missing any measures contained in corresponding data files
- T7: test_measure_info_keys
    - Checks whether measure_info files have valid keys for each variable
- T8: test_jsons
    - Checks whether encountered JSONS are valid JSONS that can be read
- T14: test_measure_info_extra_measures
    - Checks whether measure_info files have any extra measures not contained in any corresponding data files

We found the following results for core metadata validity and accuracy (As of 8/13/2023):

- T3: test_measure_info_structure
  - 141/234 (60.3%) valid
  - 93/234 (39.8%) invalid
- T4: test_measure_info_missing_measures
  - 158/237 (66.7%) valid
  - 34/237 (14.3%) missing measures
  - 45/237 (19.0%) no measure_info files
- T7: (test_measure_info_keys)
  - 749/856 (87.5%) valid
  - 107/856 (12.5%) missing keys
- T8: (test_jsons)
  - 91/91 (100%) valid
- T14: (test_measure_info_extra_measures)
  - 112/115 (97.4%) valid
  - 1/115 (0.8%) error
  - 2/115 (1.7%) extra measures

We found that poor performance on some tests was caused by outdated test specification. For example, T3 had not been updated to reflect the sub-elements that are unique to reference metadata entries or dynamic metadata, thus implying poor adherence to the proper metadata structure. Similarly, T4 showed poor performance for the geographies data repository. We have not required the presence of metadata info files in the geographies data repository and their presence is not necessary for the Data Commons architecture. This test will be refined to only include repositories where this standard is expected. Future work includes updating tests to better reflect active metadata standards.

The results of T4 and T7 illustrated of areas to improve metadata and standards adherence. Some datasets, because they are not shown on a data dashboard, were not a priority for documentation. The metadata team, though, enforced the documentation of all datasets regardless of their intention to be displayed. Thus, the areas identified by T4 as missing metadata for some or all measures signified an area of opportunity for improving metadata. Another illuminating test is T7. As previously mentioned, core metadata elements are not required. T7 shows how this standard has been interpreted and executed differently by different data producers: to omit elements entirely from a metadata record or leave them blank. This illuminated a communication breakdown on the part of the metadata team. We plan to better communicate how we expect metadata to adhere to this standard.

T8 demonstrated our progress towards the FAIR principle of interoperability. This test most directly relates to FAIR guideline I1. The results of this test showed that we are successfully implementing an interoperable framework by storing our metadata in valid JSONs. Although our local core element standard is not widely interoperable yet, we are prioritizing maintaining this standard within a more universally accessible file format.

T14 showed that a few datasets have metadata for measures that do not exist in the data. A visual inspection revealed these extra measures are older versions of current measures. In these cases, the data producer had updated their data production process and metadata but continued to produce an older version of the dataset as well. This behavior addresses, in part, A2 of the FAIR guidelines. Future work for the metadata team includes codifying a system-wide strategy to adhere to this principle. It is unclear if the nature of a version control system like GitHub, which we already adopt, provides sufficient accessibility for retired versions of data and metadata [8].

## 3. Future work

The Data Commons is growing to accommodate more external collaborators (i.e. dataset producers external to the University of Virginia) and more stakeholders, increasing the need for a robust metadata management system. External collaborators have included off-site student researchers, as well as researchers from stakeholder organizations. Having our standards codified and accessible is necessary for collaboration and adherence. Moreover, the simpler we make it for researchers to adhere to standards, such as by building tools and automatic processes for feedback, the easier it is to coordinate work with a distributed network of collaborators. More stakeholders also means an increasing diversification of stakeholder needs. Accurate and rich metadata is necessary to meet the needs of stakeholders, including the ability to access and interpret metadata to use datasets and data products.

We plan to perform user testing for the Data Commons project to evaluate metadata richness and accuracy for external users. Thus far, we have performed occasional user testing focused on system architecture and overall usability of the data dashboard. We have yet to explore user testing for topics such as accurate data interpretation using the Social Impact Data Commons. Testing if users can accurately interpret data using our descriptions, which are contained in the core metadata, will impact our process and guidelines for writing metadata.

We also plan to develop a crosswalk of our standard to the Data Documentation Initiative (DDI) to facilitate adherence to the FAIR data principles. One of the major shortcomings of our metadata solution is interoperability. While JSON itself is an interoperable metadata management standard, we have chosen mostly locally-defined core metadata elements, borrowing little from more widely adopted metadata standards. To increase the interoperability of our metadata, we aim to develop a crosswalk of our standard to DDI as well as other relevant standards, such as DataCite, Data Catalog Vocabulary (DCAT), and schema.org. We plan to develop the crosswalk so that DDI metadata is automatically generated when a data producer creates metadata based on our local standard. We could also explore developing tools so that stakeholders themselves could crosswalk the data for use in their applications.

This crosswalk could also be leveraged to automatically push out datasets to external data repositories. We hope to make our datasets more findable and accessible by including them in external data repositories. Already we have developed pipelines to push our datasets to University of Virginia's LibraData, which is our institutional instance of the Dataverse. We have leveraged the Dataverse GitHub Action to automatically push datasets up. We aim to include our datasets in other relevant repositories, such as Inter-university Consortium for Political and Social Research.

Having our datasets on external repositories will make our datasets more discoverable, achieving better adherence to FAIR. A major drawback of hosting datasets on GitHub is that GitHub is not designed or optimized to store datasets. While many of the advantages of GitHub transfer to data storage, some do not. For example, GitHub's internal search and external discoverability is optimized for findability of code. Because code, data, and metadata are stored and look very differently, GitHub's code search is not optimized for discoverability of datasets [10]. We have not noticed substantial traffic to our repositories on GitHub, nor the Dataverse. We hope that developing automatic systems to push out datasets to institutional and subject-specific repositories will increase their discoverability, and therefore their findability.

## 4. Conclusion

Metadata is an integral piece of the Data Commons. Actionable and evaluable metadata have made the process of creating the Data Commons much more efficient, as more adherent to FAIR guidelines. Core metadata, specifically, is used to build data products, supports the dissemination of statistical products, and reduces documentation burden on researchers. We are progressing our adherence to FAIR standards, with the biggest area for improvement being the creation of interoperable metadata. By evaluating our testing system, we also can improve the dissemination of standards to data producers. We also identified ways to improve the testing system itself.

Before the start of the Data Commons project, the team did not include anyone with metadata expertise. We have improved the metadata literacy of metadata team members which has benefitted all researchers in our lab. In addition to literacy, we hope to have instilled an appreciation for metadata within researchers, especially those whose discipline-specific background did not emphasize the importance of metadata. We are excited to tackle the next steps, such as improving interoperability and distributing datasets on external repositories to improve their findability.

**Table A1**

**Indicators associated with the FAIR Data Maturity Model, reproduced from FAIR Data Maturity Model Working Group (2020)** [9]

| Principle | Indicator ID | Indicators | Priority |
|---|---|---|---|
| F1 | RDA-F1-01M | Metadata is identified by a persistent identifier | Essential |
| F1 | RDA-F1-01D | Data is identified by a persistent identifier | Essential |
| F1 | RDA-F1-02M | Metadata is identified by a globally unique identifier | Essential |
| F1 | RDA-F1-02D | Data is identified by a globally unique identifier | Essential |
| F2 | RDA-F2-01M | Rich metadata is provided to allow discovery | Essential |
| F3 | RDA-F3-01M | Metadata includes the identifier for the data | Essential |
| F4 | RDA-F4-01M | Metadata is offered in such a way that it can be harvested and indexed | Essential |
| A1 | RDA-A1-01M | Metadata contains information to enable the user to get access to the data | Important |
| A1 | RDA-A1-02M | Metadata can be accessed manually (i.e. with human intervention) | Essential |
| A1 | RDA-A1-02D | Data can be accessed manually (i.e. with human intervention) | Essential |
| A1 | RDA-A1-03M | Metadata identifier resolves to a metadata record | Essential |
| A1 | RDA-A1-03D | Data identifier resolves to a digital object | Essential |
| A1 | RDA-A1-04M | Metadata is accessed through standardised protocol | Essential |
| A1 | RDA-A1-04D | Data is accessible through standardised protocol | Essential |
| A1 | RDA-A1-05D | Data can be accessed automatically (i.e. by a computer program) | Important |
| A1.1 | RDA-A1.1-01M | Metadata is accessible through a free access protocol | Essential |
| A1.1 | RDA-A1.1-01D | Data is accessible through a free access protocol | Important |

| Principle | Indicator ID | Indicators | Priority |
| --- | --- | --- | --- |
| A1.2 | RDA-A1.2-01D | Data is accessible through an access protocol that supports authentication and authorisation | Useful |
| A2 | RDA-A2-01M | Metadata is guaranteed to remain available after data is no longer available | Essential |
| I1 | RDA-I1-01M | Metadata uses knowledge representation expressed in standardised format | Important |
| I1 | RDA-I1-01D | Data uses knowledge representation expressed in standardised format | Important |
| I1 | RDA-I1-02M | Metadata uses machine-understandable knowledge representation | Important |
| I1 | RDA-I1-02D | Data uses machine-understandable knowledge representation | Important |
| I2 | RDA-I2-01M | Metadata uses FAIR-compliant vocabularies | Important |
| I2 | RDA-I2-01D | Data uses FAIR-compliant vocabularies | Useful |
| I3 | RDA-I3-01M | Metadata includes references to other metadata | Important |
| I3 | RDA-I3-01D | Data includes references to other data | Useful |
| I3 | RDA-I3-02M | Metadata includes references to other data | Useful |
| I3 | RDA-I3-02D | Data includes qualified references to other data | Useful |
| I3 | RDA-I3-03M | Metadata includes qualified references to other metadata | Important |
| I3 | RDA-I3-04M | Metadata include qualified references to other data | Useful |
| R1 | RDA-R1-01M | Plurality of accurate and relevant attributes are provided to allow reuse | Essential |
| R1.1 | RDA-R1.1-01M | Metadata includes information about the licence under which the data can be reused | Essential |
| R1.1 | RDA-R1.1-02M | Metadata refers to a standard reuse licence | Important |
| R1.1 | RDA-R1.1-03M | Metadata refers to a machine-understandable reuse licence | Important |
| R1.2 | RDA-R1.2-01M | Metadata includes provenance information according to community-specific standards | Important |
| R1.2 | RDA-R1.2-02M | Metadata includes provenance information according to a cross-community language | Useful |
| R1.3 | RDA-R1.3-01M | Metadata complies with a community standard | Essential |
| R1.3 | RDA-R1.3-01D | Data complies with a community standard | Essential |
| R1.3 | RDA-R1.3-02M | Metadata is expressed in compliance with a machine-understandable community standard | Essential |
| R1.3 | RDA-R1.3-02D | Data is expressed in compliance with a machine-understandable community standard | Important |